\documentclass[12pt]{iopart}

\newcommand{\egy}[1]{\begin{equation}#1\end{equation}}
\newcommand{\dd}{\mathrm{d}}
\usepackage{iopams}
\usepackage{setstack}
\usepackage{graphics}
\usepackage{graphicx}

\begin{document}

\title[Fluctuations in the zero temperature XX chain: Effective temperature]
{Fluctuations in subsystems of the zero temperature XX chain:
Emergence of an effective temperature}

\author{V. Eisler$^1$, \"O. Legeza$^2$ and Z. R\'acz$^1$}
\address{$^1$ Institute for Theoretical Physics,
E\"otv\"os University, 1117 Budapest, P\'azm\'any s\'et\'any 1/a, Hungary}
\address{$^2$ Research Institute for Solid State Physics and Optics, H-1525
Budapest, P.\ O.\ Box 49, Hungary}
\eads{\mailto{eisler@general.elte.hu}, \mailto{olegeza@szfki.hu} and
\mailto{racz@general.elte.hu}}

\begin{abstract}
The zero-temperature XX chain is studied with emphasis
on the properties of a block of $L$ spins inside the chain.
We investigate the quantum fluctuations resulting from the
entanglement of the block with the rest of the chain
using analytical as well as numerical (density matrix
renormalization group) methods. It is found that
the rest of the chain acts as a thermal environment and an
effective temperature can be introduced to describe the
fluctuations. We show that the effective temperature
description is robust in the sense that several independent
definitions (through fluctuation dissipation theorem,
comparing with a finite temperature system) yield the same
functional form in the limit of large block size ($L\to\infty$).
The effective temperature can also be shown to satisfy
the basic requirements on how it changes when two bodies of
equal or unequal temperatures are brought into contact.
\end{abstract}

\section{Introduction}
Entanglement properties of various spin chains have attracted much
attention recently. The most intensively investigated cases are spin chains in
their zero temperature ground states, where the entanglement between a
contiguous block of spins and the rest of the chain is measured through
the von Neumann entropy. Thus far, the entanglement entropy has been
found to be a well behaving measure of entanglement which shows some universal
features at quantum critical points
\cite{Vidal,Korepin,Keating,CalabCardy,Eisler}. Despite capturing some
important aspects, one should not forget, that we inevitably lose
information on the entangled state by using only
a coarse grained measure. Indeed, the restriction of the pure ground
state to a subsystem results in a mixed state with quantum fluctuations,
and the weights are given by the full spectrum of the reduced density matrix,
which means a number of parameters proportional to the exponential of
the system size.
Hence an ``integrated'' measure such as the entropy will not tell about
the distribution of the fluctuations. This gives an obvious motivation
for a deeper analysis of reduced density matrix spectra.
\par
In the past few years, much effort has been devoted to the
investigation of reduced density matrices in a series of simple
model systems \cite{Legeza,Peschelosc,Peschel2D,Peschelferm,Orus}.
The results obtained indicate that, in several cases, the density
matrix itself can be expressed as the exponential of a quadratic
Hamiltonian. An example is the case of free fermions in one dimension
(equivalent to an XX chain) which was investigated in detail
\cite{PeschelXXlett,Henley,PeschelXX} and the reduced density
matrix for arbitrary subsystem sizes has been obtained in a closed form.
\par
Here we try to develop a simple interpretation of the above results,
concentrating on the XX chain. Looking at a subsystem of $L$
spins in an infinitely large chain, the quantum fluctuations in the block
are due to the coupling with the rest of the chain. One could think of an
analogy with the subsystem-reservoir scheme of the canonical ensemble
in statistical physics where now the spins outside the block would play
the role of the reservoir. Therefore it is an intriguing question whether
it is possible to define an ``effective temperature'' to describe the zero
temperature quantum fluctuations with Boltzmann-like distributions, 
meaning that the weights of the reduced density matrix $\hat\rho_L$ decay
exponentially with the subsystem energies:
\egy{\hat\rho_L = \frac 1 Z e^{-\beta_{eff}\hat H_L}\label{rhogibbs},}
where $\hat H_L$ represents the Hamiltonian of the subsystem,
and $\beta_{eff}$ denotes the effective inverse temperature.
The aim of this paper is to show that there are several ways to
define an effective (inverse) temperature but they are all consistent
in the sense that the leading order dependence on
the subsystem size is identical, given by:
\egy{\beta_{eff}J=\pi \frac{L}{\ln L},}
where $J$ is the coupling between nearest-neighbour spins.
In the following we shall present a detailed derivation
of the above result as outlined below.
\par
In Section \ref{sec:RDM}, we give a brief summary of the
results on the reduced density matrix of the XX chain
\cite{PeschelXXlett}. Next (Sec.~\ref{sec:1pspect}),
we analyze the one-particle spectra of the effective Hamiltonian
emerging in the formula of the density matrix and that of the Hamiltonian
describing a finite XX chain. We define the effective temperature
starting out from a continuum limit formula for the eigenvalues
\cite{PeschelXX}, then verify these results by analyzing the multiparticle
spectra of finite systems either by means of the operator based truncation
scheme \cite{Henley} or density matrix renormalization group (DMRG)
calculations \cite{White}.
\par
Section \ref{sec:fluctdis} is concerned with an alternative
definition of the effective temperature, using an extension
of the fluctuation dissipation theorem to zero 
temperature. The calculations are carried through for two
different conserved quantities and the results obtained
show agreement with the previous definitions.
\par
With the concept of a system in an effectively thermodynamic 
environment, we try a direct mapping in Section \ref{sec:map}.
The effective temperature is determined numerically by
comparing canonical and zero temperature expectation values.
Surprisingly, we recover the results of the fluctuation-dissipation
ratio with a very high precision, and an explanation for the precise
agreement is also given.
\par
The definitions are also tested by studying their consistency
with thermodynamics (Sec.~\ref{sec:contact}). Namely, we ask
what happens when subsystems of equal or different temperatures
are brought into contact. We find that equal temperatures do not change,
while in case of unequal temperatures the resulting temperature is
in between the initial ones.
\par
Finally, we summarize our results (Sec.~\ref{sec:fin})
emphasizing also the problematic points which might
set a limit on the applicability of the simple picture described here.
\section{Reduced density matrix for the XX chain \label{sec:RDM}}
Our starting point for the calculations is the Hamiltonian
of the XX chain
\egy{\hat H^{XX}=-J\sum_{j=1}^{N-1}{(s_j^x s_{j+1}^x +
s_j^y s_{j+1}^y)}- h \sum_{j=1}^N{s_j^z},\label{xxham}}
where $s^\alpha_j (\alpha=x,y,z)$ are the Pauli spin matrices at
sites $j=1,2,\dots,N$ of the chain, $J$ is the coupling and $h$
is the magnetic field. All our results, unless stated otherwise,
are derived in the case of zero magnetic field ($h=0$).
It is well known that this model can be transformed into a chain
of free Fermions \cite{LSM} with Hamiltonian
\egy{\hat H = -\frac 1 2 \sum_{j=1}^{N-1}
( c_j^\dag c_{j+1} + c_{j+1}^\dag c_j ),\label{ffham}}
where $c_j^\dag$ and $c_j$ are fermionic creation and annihilation
operators and energy is measured in units of the coupling $J$.
\par
If one considers only a subsystem of size $L$ in a large chain
with $N$ spins, the state of the subsystem is fully described
by its reduced density matrix 
$\rho_L=\mathrm{tr}_{N-L}|\Psi_0\rangle\langle \Psi_0 |$, 
obtained from the ground state $|\Psi_0\rangle$ by tracing out external
degrees of freedom. It was previously shown \cite{PeschelXXlett} that
for a chain with Hamiltonian (\ref{ffham}) the reduced density
matrix can be written in an exponential form
\egy{\hat\rho_L = \frac{1}{\tilde Z} e^{-\tilde H} \label{rhol}}
where $\tilde Z=\mathrm{Tr}(e^{-\tilde H})$ is a normalization constant
and $\tilde H$ is an effective Hamiltonian of the following form:
\egy{\tilde H = \sum_{i,j} A_{ij}c^\dag_i c_j.}
One can see that $\tilde H$ is quadratic in the fermionic operators.
The matrix elements $A_{ij}$ can be given through the elements of the
correlation matrix $C_{ij}=\langle c_i^\dag c_j\rangle$ 
by the following relation:
\egy{A=\ln\frac{1-C}{C},}
where $A$ and $C$ denote the matrices composed of the elements
$A_{ij}$ and $C_{ij}$, respectively.
Hence, the effective Hamiltonian can be diagonalized, 
and the eigenvalues can be obtained through the eigenvalues $\xi_k$
of the correlation matrix:
\egy{\tilde H = \sum_{k=1}^L \varepsilon_k f^\dag_k f_k \, ,\quad
\varepsilon_k=\ln\frac{1-\xi_k}{\xi_k},\label{effhamspect}}
where $f^\dag_k$ and $f_k$ are fermionic operators
obtained by unitary transformation.
\par
The diagonalization of the correlation matrix is especially simple
if we take the whole system to be infinitely large ($N\to\infty$).
In this case the elements of the matrix depend only on the distance
$r=|i-j|$ between the sites and can be written as:
\egy{C_{ij}=\frac{\sin(\frac \pi 2(i-j))}{\pi(i-j)}=
\left\{\begin{array}{lr}
\frac{(-1)^{\frac {r-1}{2}}}{\pi r}, & r\, \mathrm{odd}\\
0, & r\, \mathrm{even}\\
\frac 1 2, & r=0
\end{array}.\right.\label{corrm}}
For finite $N$, each element of the correlation matrix has
to be evaluated numerically as a sum of $N$ terms.
\par
The purpose of studying the structure of the reduced density matrix
$\hat\rho_L$ came from our intention to define an effective temperature
by making the correspondence between the density matrix weights
and Boltzmann factors (\ref{rhogibbs}) with respect to the
subsystem Hamiltonian $\hat H_L$. This latter one can be diagonalized
exactly and it also transforms to a system of noninteracting fermions:
\egy{\hat H_L = \sum_{k=1}^L \lambda_k c^\dag_k c_k \, ,\quad
\lambda_k=-\cos\left(\frac{\pi k}{L+1}\right).\label{hlspect}}
\par
Peschel's results show that
$\hat\rho_L$ can be written as an exponential of a free fermion 
Hamiltonian (\ref{effhamspect}). Thus, from comparison with (\ref{hlspect})
it can be seen, that all the information we need is stored in the 
one-particle spectra $\varepsilon_k$ and $\lambda_k$, where the former
one has to be numerically calculated.
\section{One-particle and multiparticle spectra \label{sec:1pspect}}
\begin{figure}[thb]
\center
\includegraphics[scale=0.55,angle=270]{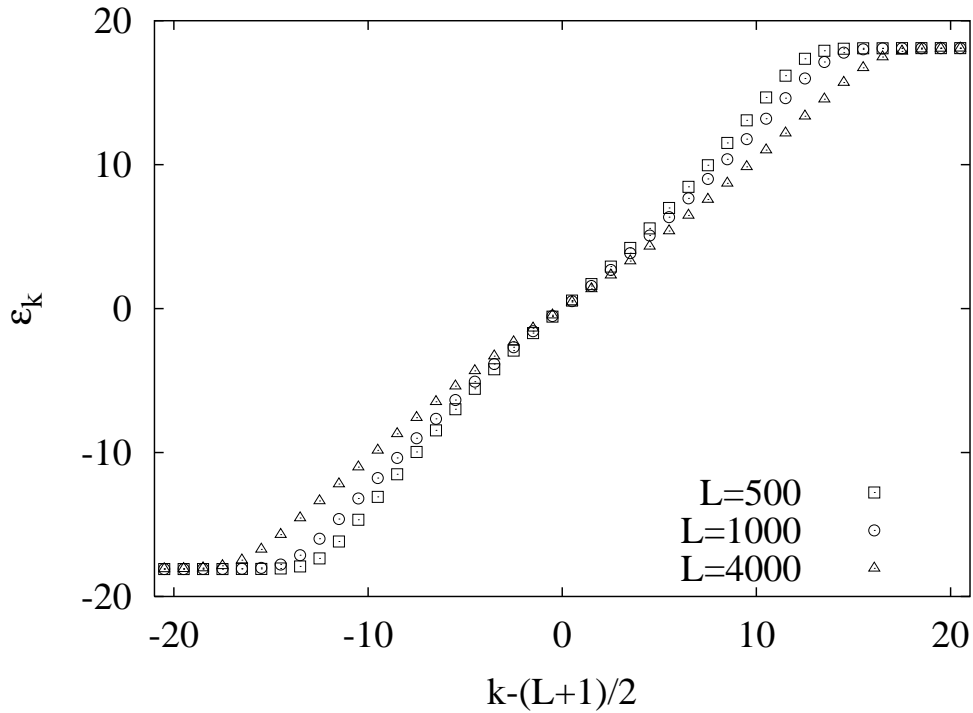}
\caption{One-particle spectra of the effective Hamiltonian $\tilde H$
for various block sizes $L$ with the chain size $N$ taken to be infinite.
The spectrum is shifted for each $L$ to visualize
the particle-hole symmetry. The saturation of the curves for
$|k-(L+1)/2|>10$ is due to numerical errors.}
\label{fig:effhamspect}
\end{figure}

Figure \ref{fig:effhamspect} shows the one-particle spectra of
the effective Hamiltonian $\tilde H$ for various subsystem sizes
with the chain size $N$ taken to be infinite.
First of all, we should notice the apparent particle-hole symmetry
of the spectra, which is made more visible by shifting the curves
to a common origin. This symmetry property is not surprising as 
it is already present in the spectrum of the XX chain with zero 
magnetic field. The saturation of the curves for $|k-(L+1)/2|>10$ 
is due to numerical errors, since the major part of the eigenvalues
of the correlation matrix $C_{ij}$ lie exponentially close to $0$ and $1$.
%
\par
The weights of the density matrix can be produced from these
one-particle spectra in the same way as in the case of a real
excitation spectrum: the largest weight is obtained by filling
the modes up to the Fermi level, i.e. all modes that are negative.
The next largest weights are obtained by filling the smallest
positive $\varepsilon_k$ (``particle excitation'') or removing
the smallest negative one (``hole excitation'').
Thus, the problem of finding the $2^L$ weights of the density matrix
can be reduced to the diagonalization of an $L \times L$ matrix.
\subsection{Continuum limit}
Taking a continuum limit in the block 
($L \to \infty$) Peschel calculated analytically the lowest lying eigenvalues
\cite{PeschelXX}:
\egy{\varepsilon_j = \pm \frac{\pi^2}{2 \ln L}(2j-1),\quad j=1,2,\dots
\label{epsasympt}}
%
One can see that the spectrum is linear near the Fermi level
and the eigenvalues scale with $1/\ln L$. Looking at the excitation 
spectrum of the block XX Hamiltonian $\hat H_L$, one finds that it is
also linear near the Fermi level and, for large block sizes, it can 
be approximated as:
\egy{\lambda_{k=L/2+j}=-\cos\left(\frac \pi 2 + \frac \pi 2
\frac{2j-1}{L+1}\right)\approx \frac{\pi}{2L}(2j-1).\label{fermilevel}}
Hence, the effective inverse temperature can be obtained by
comparing only the low lying part of the one-particle spectra.
From Eq. (\ref{epsasympt}) and (\ref{fermilevel}) one has:
\egy{\beta_{eff}=\pi \frac{L}{\ln L}.\label{eq:betaeff}}
It can be seen that, according to the expectations,
$\beta_{eff} \to \infty$ when $L \to \infty$, that is
we recover the zero temperature ground state if we take
the whole chain as the subsystem. It has to be emphasized
that the above definition of the effective
inverse temperature describes correctly only the low energy part
of the spectrum. However, it is well known from DMRG calculations
that indeed, it is only a small portion of the weights which are
relevant for the physics of the system \cite{White}, therefore we
also expect that the effective temperature is reasonably defined.
\par
We should note that, comparing with numerical results, one finds 
deviations from the asymptotic formula (\ref{eq:betaeff}).
In the paper of Cheong and Henley \cite{Henley}, where
the idea of defining an effective temperature has first appeared,
the logarithmic scaling of the eigenvalues could not be detected
for system sizes $L<100$, hence they found that $\beta_{eff}\propto L$.
This can be understood since the logarithmic dependence on the
block size makes convergence very slow and one needs corrections
even for large block sizes ($L=2000-5000$).
Determining $\beta_{eff}$ as the ratio of the lowest eigenvalues
of (\ref{epsasympt}) and (\ref{fermilevel})
we have tried a fit $a \times L/(\ln L +b)$ which gave values 
$a=3.12 \approx \pi$ and $b=2.58$ which shows 
considerable finite size corrections, already noted in \cite{PeschelXX}.
\subsection{Operator-based truncation and DMRG results}
In order to compare the results of the continuum limit for the
effective temperature (\ref{eq:betaeff}) to the one defined in 
Eq.(\ref{rhogibbs}), one has to consider finite
size systems and build the multiparticle spectra from the
one-particle spectra. A possible method is the so-called
operator-based truncation scheme which was devised by
Cheong and Henley \cite{Henley} and consists of truncating
the one-particle spectrum by retaining only a small number
of the excitations symmetrically around the Fermi level.
The method can be applied successfully, since the normalization
constant $\tilde Z$ itself can be determined by using only the
retained eigenvalues. Indeed, it can be verified \cite{Henley} that
\egy{\frac{1}{\tilde Z}= \det (1-C)=\prod_{k=1}^{L}\frac{1}
{1+e^{-\varepsilon_k}}.}
For large positive eigenvalues $1+e^{-\varepsilon_k} \approx 1$
while for large negative ones 
$1+e^{-\varepsilon_k}\approx e^{-\varepsilon_k}$ thus the latter
terms simply cancel the factors of the occupied one-particle states
in the exponential (\ref{rhol}).
Hence the most relevant weights of the reduced density matrix can be
reproduced from the low lying excitations of the one-particle spectrum and
the highly excited states, corrupted by numerical error, can be discarded.
\par
The same method can be applied to obtain the multiparticle
energy eigenvalues of the block, except that in this case
we can calculate the given subset of the spectrum exactly,
since the one-particle energies are known analytically.
With this systematic procedure a one-to-one correspondence
between the block energies and the density matrix weights
can be established.
\par
The multiparticle spectra constructed from the one-particle spectra 
has also been confirmed by the DMRG method \cite{White}.
In the DMRG method, the finite system is divided into two subsystems,
thus the multiparticle spectra $\omega_\alpha(L)$ of the reduced 
subsystem density matrix and the eigenvalues $E_\alpha(L)$ of the reduced
subsystem Hamiltonian can be calculated directly for a
block of $L$ spins, where $\alpha$ runs from one to the number of basis
states, $M$, used to describe the block.
We have calculated $\omega_\alpha(L)$ and $E_\alpha(L)$ 
as a function of the subsystem size $L$ on finite chains with $N$
lattice sites with open boundary condition using the dynamic 
block-state selection (DBSS) approach \cite{legeza03}. 
The DBSS method allows a more rigorous control of the numerical accuracy
by setting the threshold value of the quantum information loss $\chi$ and
the minimum number of basis states, $M_{\rm min}$, used to describe the 
subsystem blocks a priory of the calculations.
For a given finite system with $N$ lattice sites subsystem blocks
with $L=2,\dots,N-2$ can be obtained using the sweeping procedure of the 
so called finite lattice algorithm. 
\par
\begin{figure}[htb]
\center
\includegraphics[scale=0.75]{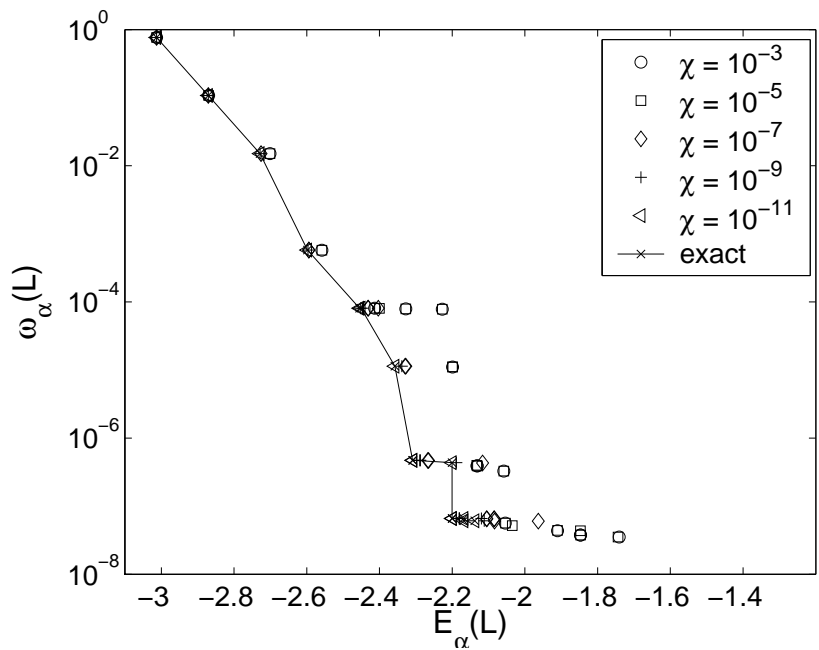}
\caption{Eigenvalue spectrum of $\hat \rho_L$ as a function of
the eigenvalues of $\hat H_L$ with $L=10$ for a chain with $N=20$
lattice sites for various $\chi$ values. The result of
the exact diagonalization is shown by the solid line.}
\label{fig:rho_chi}
\end{figure}

First, we have compared exact results obtained by exact diagonalization 
for $L=N/2$ on a short $N=20$ chain to the DMRG results by systematically
adjusting $\chi$.
In Fig.~\ref{fig:rho_chi} we plotted $\omega_\alpha(L)$ as a function of
$E_\alpha(L)$ obtained after the sixth DMRG sweep for various
$\chi$ values with $M_{\rm min}=16$.
The accuracy of the diagonalization 
of the superblock Hamiltonian was set to $10^{-9}$.   
%
%
The agreement between the exact diagonalization and DMRG results is evident
from the figure for large $\omega_\alpha(L)$ even for large $\chi$ values.
On the other hand, for the small $\omega_\alpha(L)$ spectrum more accurate
calculations with smaller $\chi$ values have to be performed for better
agreement.
\par
\begin{figure}[htb]
\center
\includegraphics[scale=0.75]{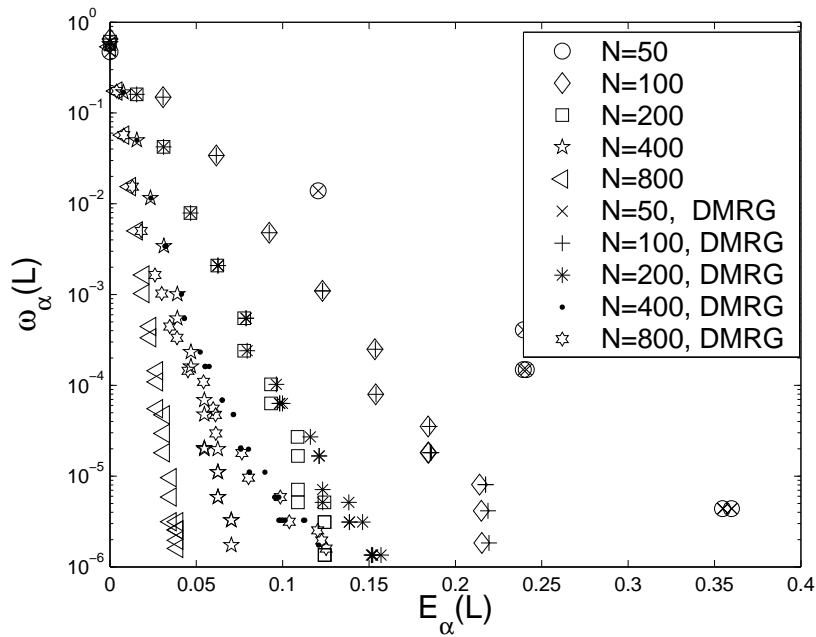}
\caption{Eigenvalue spectrum of $\hat \rho_L$ as a function of the
eigenvalues of $\hat H_L$ with $L=N/2$ for $N=50,100,200,400,800$
using the operator-based truncation method and the DMRG algorithm with 
six sweeps. Degeneracies in the spectra are removed and all $E_\alpha(L)$ are 
measured from zero.}.
\label{fig:rho_n_w2048}
\end{figure}
\begin{figure}[htb]
\center
\includegraphics[scale=0.75]{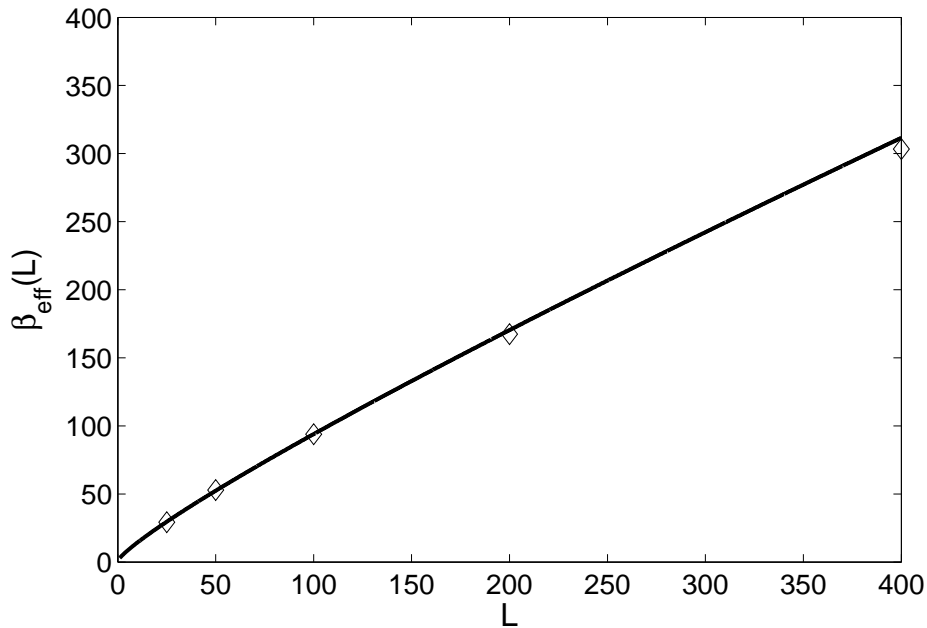}
\caption{$\beta_{eff}(L)$ as function of $L=N/2$ obtained as the absolute
value of the slopes of straight line fits to the $\ln (\omega_\alpha(L))$ vs.
$E_\alpha(L)$ data sets. The solid line is the result of the fit with
$a=6.309$ and $b=2.101$.}
\label{fig:beta_l_obt}
\end{figure}

In order to determine the size dependence of the effective temperature 
similar calculations were repeated first for subsystems with $L=N/2$
by varying $N$ and then by keeping the size of the system fixed 
but varying the size of the block. For the former case
Fig.~\ref{fig:rho_n_w2048} shows our results for $N=50,100,200,400,800$
using DMRG method with six sweeps, $\chi=10^{-8}$, $M_{\rm min}=1024$
and the operator-based truncation method with 20 particle-hole
excitations kept.
%
%
For large $\omega_\alpha (L)$ eigenvalues the agreement between the
operator-based truncation and DMRG result is evident from the figure,
while for the lower part of the spectrum 
the deviation between the two approaches due to truncation procedures is
more significant. 
This result, however, confirms the validity of the
operator-based truncation method and it is sufficient in order to determine
the effective temperature. 
Data points corresponding to $\omega_\alpha(L)>10^{-4}$ eigenvalues 
for each $L$ can be fitted with a straight line in order to obtain
$\beta_{eff}(L)$ which is shown in Fig.~\ref{fig:beta_l_obt}. 
Our result indicates that $\beta_{eff}(L)$ can be fitted in leading order
with the  following ansatz $a \times L/(\ln L+b)$ and we obtained 
$6.42 > a > 5.76$ and $2.12 > b > 1.57$ depending on the number of
data points taken into account when $\beta_{eff}(L)$ was fitted by
the straight lines. Using the operator-based truncation approach we
repeated the same calculations with $L=N/4$ for $N=200,400,800,1600,3200$
and found again that $a$ changed between $6.37$ and $5.63$.
The value of $a$ close to $2\pi$ is twice that of given in 
Eq.~(\ref{eq:betaeff}). This is due to the fact that the blocks were
always determined for a semi-infinite system having one open boundary.
When the same calculations are carried out for chains with up to $N=200$
lattice sites using DMRG with periodic boundary condition (PBC) in which
case there are two couplings across the block interface we obtained
$3.3 > a > 2.9$ and $ 2.1 > b > 1.8$ in agreement with 
Eq.~(\ref{eq:betaeff}).
%
%
\par
\begin{figure}[htb]
\center
\includegraphics[scale=0.75]{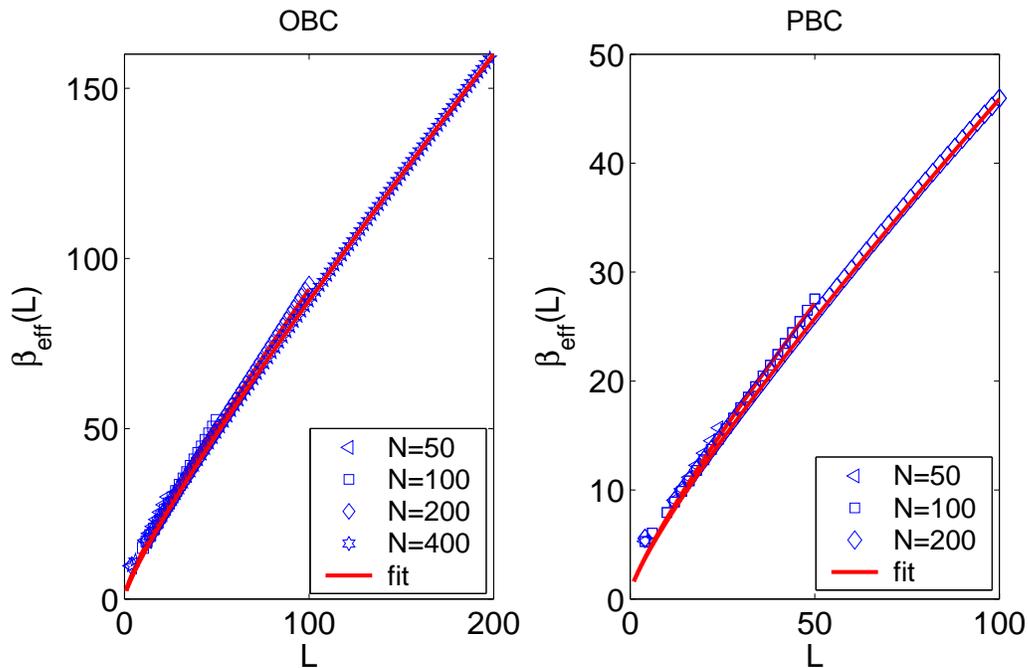}
\caption{$\beta_{eff}(L)$ as function of $L$ obtained as the 
absolute value of the slopes of straight line fits to the 
$\ln (\omega_\alpha(L))$ vs. $E_\alpha(L)$ data sets
corresponding to finite chains with $N=50,100,200,400$ lattice sites.
DMRG results, shown in the left and right panel, were obtained with open
and periodic boundary condition, respectively. The solid lines are the
results of the fitting.}
\label{fig:beta_l}
\end{figure}

We have also calculated $\omega_\alpha(L)$ and $E_\alpha(L)$ for 
various finite chains with $N=50,100,200,400$ lattice sites for DMRG 
blocks with $L=2,...,N/2$ using several sweeps of the finite-lattice algorithm.
For a given $N$ taking data set corresponding to $\omega_\alpha(L)>10^{-4}$
a linear fit for each $L$ can be obtained in order to determine 
$\beta_{eff}(L)$ as a function of $L$. Our result shown in the left panel
of Fig.\ref{fig:beta_l} indicates that in leading order
$\beta_{eff}(L)$ can again be fitted very well with the following ansatz
$a \times L/(\ln L+b)$. For all fits we obtained $6.38 > a > 6.21$
and $b$ changed between $2.08$ and $2.49$. We also found that these fits
are more stable than the above described previous approach.
%
%
The value of $a$ is again very close to $2\pi$.  
When the same calculation is repeated using PBC we found $3.28 > a > 2.94$
and $2.12 > b > 1.81$ in agreement with Eq.~(\ref{eq:betaeff}) as also shown
in the right panel of Fig.\ref{fig:beta_l}. The factor of two in $a$
for OBC and PBC is the same effect that has been discussed for the entropy 
profile of a finite subsystem for critical models with open or periodic
boundary conditions \cite{CalabCardy}. Finally, we conclude that a straight
line fit is not enough to describe $\beta_{eff}(L)$ as a function of $L$,
but logarithmic corrections still could not be confirmed rigorously due
to the limited system sizes.
\section{Fluctuation-dissipation ratio \label{sec:fluctdis}}
To support the idea of an effectively thermodynamic system,
it would be desirable to find some alternative way of defining 
a temperature, and to compare the results of the different approaches.
In ordinary thermodynamic systems, coupled to a reservoir at some
finite temperature, the fluctuation-dissipation theorem gives
a possible way of defining the temperature. If we restrict ourselves
to the static limit and consider only conserved quantities of the Hamiltonian,
it reads:
\egy{\chi_{A}=\beta (\langle \hat A^2 \rangle_\beta
-\langle \hat A \rangle_\beta^2) ,}
where $\hat A$ is a conserved quantity, and 
$\chi_A=\frac{\partial\langle \hat A \rangle_\beta}{\partial h_A}$
is the response function of $\hat A$ to an infinitesimal conjugate field
$h_A$. The brackets denote averages taken with the canonical ensemble.
\par
The straightforward generalization to our case would be to define the
effective inverse temperature as the ratio of the response function
to the fluctuations inside the block. We shall present here the
calculations for two cases, using the transverse
magnetization or the magnetization flux as the observable $\hat A$.
From now on we shall consider only the case when the chain size
$N$ is taken to be infinite.
\subsection{Fluctuations of the transverse magnetization \label{sec:magfluct}}
The transverse magnetization is a conserved quantity in the XX model,
thus it is a good candidate for the definition of the effective
temperature, as described above:
\egy{\beta_{eff}=\frac{\chi_M}
{\langle \hat M_L^2 \rangle-\langle \hat M_L \rangle^2}
\label{betafluctdis}}
where $\chi_M$ is the magnetic susceptibility, $\hat M_L$ is the transverse
magnetization of the block, and the brackets now denote averages
taken with $\hat\rho_L$. It is important to observe that we do not
need the reduced density matrix to evaluate the expectation values
since it is equivalent to calculating them in the ground state 
$|\Psi_0\rangle$ of the whole chain, which is known exactly.
\par
Let us begin by calculating the block magnetization in case of an applied
magnetic field. The ground state $| \Psi_h \rangle$ is known exactly 
\cite{LSM}, and we have:
\egy{
\langle \Psi_h| \hat M_L | \Psi_h \rangle = 
\sum_{j=1}^{L}\langle c_j^\dag c_j -\frac 1 2 \rangle 
=\frac{L}{2\pi} \int_{-\pi}^{\pi} \langle c_q^\dag c_q\rangle \dd q- \frac L 2 
= L(\frac{q_h}{\pi}-\frac 1 2),\label{eq:maghinf}}
where $q_h=\arccos (-h)$ is the Fermi wave number. The magnetic 
susceptibility is therefore
\egy{\chi_M=\left. \frac{\partial \langle \hat M_L \rangle}{\partial h}
\right|_{h=0}=\frac L \pi,\label{magsusc}}
The fluctuations in (\ref{betafluctdis}) have to be evaluated at $h=0$, 
when the block magnetization $\langle \hat M_L \rangle$ is vanishing, 
hence the fluctuations read:
\egy{\langle \hat M_L^2\rangle= \sum_{i,j=l}^{L}\langle s_i^z s_{j}^z\rangle=
\sum_{i,j=1}^{L} \langle c_i^\dag c_i c_j^\dag c_j\rangle -\frac{L^2}{4}.}
Using Wick's theorem and introducing the difference $r=i-j$
of the sites we obtain
\egy{
\langle \hat M_L^2\rangle = -2\sum_{i < j=2}^{L} C_{ij}^2+\frac L 4 =
-\frac{2}{\pi^2} \sum_{\substack{r=1 \cr r\,\mathrm{odd}}}^{L-1}
\frac{L-r}{r^2} + \frac L 4,\label{eq:magfluctex}}
where the elements of the correlation matrix $C_{ij}$ were given
in Eq. (\ref{corrm}). It is easy to see that the linear terms in $L$
cancel, and the asymptotic form of the fluctuations will be:
\egy{\langle \hat M_L^2\rangle\approx \frac{1}{\pi^2}(\ln L + c)\, ,\quad
c= \ln 2 +\gamma+1.\label{eq:magfluct}}
Note, that (\ref{eq:magfluct}) is consistent with a result in \cite{Gioev}
where the leading order size dependence of the particle number 
fluctuations has been obtained in case of free gapless fermions in arbitrary
dimensions. Therefore the effective inverse temperature is obtained as:
\egy{\beta_{eff} = \pi \frac{L}{\ln L+c},\label{betaeff2}}
which is a form very similar to (\ref{eq:betaeff}), however there is
a correction to the logarithm already seen at the numerical
evaluation of the one-particle spectra.
\subsection{Fluctuations of the magnetization current}
To further verify the validity of our description, we shall
calculate the fluc\-tu\-a\-tion-dissipation ratio for an other
conserved quantity, the magnetization current. When the whole system
is in equilibrium, the expectation value of the current flowing
through the subsystem is vanishing, though there are still current fluctuations
inside the block, which can be calculated using the equilibrium ground state.
\par
However, in order to obtain the response function, one needs to drive
the system out of equilibrium. Therefore we need the ground state of
the chain with a magnetization flux, a problem which has been solved
by Antal \etal in \cite{xxwc}. We briefly summarize the method leading
to the main results and refer to \cite{xxwc} for the details.
\par
First, one needs to calculate the current density
\egy{\hat j_l=s^y_l s^x_{l+1}-s^x_l s^y_{l+1}=
\frac {1}{2i}(c^\dag_l c_{l+1} - c^\dag_{l+1} c_l),}
and then introduce a modified Hamiltonian $\hat H^M= \hat H - \lambda \hat J$
with the Lagrange-multiplier $\lambda$, where $\hat J=\sum_l \hat j_l$.
Note, that $\lambda$ will now play the role of the conjugate field
of the magnetization current.
\par
The ground state of the above modified Hamiltonian will give us a current
carrying steady state of the XX chain with a prescribed amount of current.
Since $[\hat H,\hat J ]=0$, Hamiltonian $\hat H^M$ can be 
diagonalized exactly and the resulting one-particle spectrum has the 
following form ($h=0$):
\egy{\Lambda_q = -\frac{1}{\cos\varphi}\cos(q-\varphi),}
with $\varphi =\arctan(\lambda)$. One can see that the
spectrum is just the original XX spectrum shifted by the
wavenumber $\varphi$, and rescaled by $\cos\varphi$.
Having obtained the structure of the ground state $|\Psi_\lambda\rangle$,
the elements of the correlation matrix can be evaluated and yield:
\egy{\langle \Psi_\lambda | c^\dag_l c_m |\Psi_\lambda\rangle = g(m-l)=
\left\{\begin{array}{lr}
\frac{(-1)^{\frac {r-1}{2}}}{\pi r} e^{ir\varphi}, & r\, \mathrm{odd}\\
0,& r\, \mathrm{even}\\
\frac 1 2, & r=0
\end{array}\right. , \quad r=m-l.}
\par
We are interested in the response function of the block to an
infinitesimal change in the conjugate field $\lambda$.
The expectation value of the magnetization current has to be
evaluated in the current carrying ground state:
\egy{\langle \Psi_\lambda | \hat J_L |\Psi_\lambda\rangle=
\frac{1}{2i}\sum_{l=1}^{L-1}\left( g(1)-g^*(1) \right)=
\frac{L-1}{\pi}\sin\varphi=\frac{L-1}{\pi}\frac{\lambda}{\sqrt{1+\lambda^2}},}
hence the response function is given by:
\egy{\left. \frac{\partial \langle \hat J_L \rangle}{\partial \lambda}
\right|_{\lambda=0}=\frac{L-1}{\pi},\label{curresp}}
which, for large values of $L$, is the same as the magnetic 
susceptibility (\ref{magsusc}).
\par
Next, the current fluctuations have to be calculated in equilibrium.
With $\langle \hat J_L \rangle=0$, we need only:
\egy{\langle \hat J^2_L \rangle =
-\frac 1 4 \sum_{l,m=1}^{L-1}\langle (c_l^\dag c_{l+1} - c_{l+1}^\dag c_l)
(c_m^\dag c_{m+1} - c_{m+1}^\dag c_m) \rangle.}
Using Wick's theorem, the contribution from terms with $|m-l|> 1$ is
\egy{-\frac 1 2 \sum_{|m-l|>1}\left[|g(m-l)|^2 - \mathrm{Re} 
\left(g(m-l+1)g^*(m-l-1)\right)\right],}
while the $|m-l|=0,1$ terms give
\egy{\frac{L-1}{2}\left[g(0)(1-g(0))+\frac{g^2(1)+g^2(-1)}{2}\right]
-(L-2)|g(1)|^2.}
Introducing $r=m-l$ and using the equilibrium value of the matrix 
elements $g(r)$, the fluctuations read:
\egy{\langle \hat J^2_L \rangle=
\frac{L-1}{2}\left(\frac 1 4 + \frac{1}{\pi^2}\right)-
\frac{1}{\pi^2}\left(\sum_{r\,\mathrm{even}}^{L-2}\frac{L-1-r}{r^2-1}+
\sum_{r\,\mathrm{odd}}^{L-1}\frac{L-1-r}{r^2}\right).}
The asymptotics of the above expression can be given as
\egy{\langle \hat J_L^2\rangle\approx \frac{1}{\pi^2}(\ln L + c^\prime)\,
,\quad c^\prime = \ln 2 +\gamma+1/2,\label{eq:currfluct}}
which differs slightly from (\ref{eq:magfluct}) in the next to
leading order term. Looking at Eqs. (\ref{curresp}) and (\ref{eq:currfluct})
we recover again the same leading order size dependence of $\beta_{eff}$.
\par
To conclude, we emphasize that the results obtained
for the effective temperatures coincide in leading order.
This is important since now contributions from all of the excited
states are included in definition (\ref{betafluctdis}),
while the theoretical construction in Section \ref{sec:1pspect} used only the
lowest lying states.
\section{Mapping to a thermalized system \label{sec:map}}
In the previous Section we defined the effective temperature
starting out from the fluctuation-dissipation theorem,
which is known to be valid for real thermodynamic systems
at finite temperature. The idea behind our approach is
a possible mapping of the subsystem in the infinite chain
at $T=0$ to a finite system which is thermalized at some
finite temperature.
\par
A more direct way of testing this mapping would be to compare
expectation values in both of the above mentioned settings. That is,
one has to consider an XX chain of finite length $L$ connected to
a heat bath, and try to adjust the temperature in such a way
that the expectation value of a given quantity would be
exactly the same as in the entangled block at zero temperature.
\par
Since the total transverse magnetization $\hat M$ of the finite
chain is vanishing at any temperature for $h=0$, we shall choose
the fluctuations to compare. Thus, the effective inverse temperature $\beta^*$
is defined from the equality
\egy{\langle \hat M^2 \rangle_{\beta^*}=\langle \hat M_L^2 \rangle
\label{betaeffnum},}
where the subscript on the bracket denotes an
expectation value taken in the canonical ensemble with
an inverse temperature $\beta^*$, while the right hand side
is the $T=0$ fluctuation. We changed the notation
of the effective inverse temperature to avoid confusion with
former definition (\ref{betafluctdis}). The r.h.s of the equation
was already determined in (\ref{eq:magfluctex}), thus one has
to evaluate only the canonical expectation value.
\par
First we calculate the partition function $Z=\mathrm{Tr}(e^{-\beta \hat H_L})$.
Since the XX Hamiltonian can be transformed into a set of
noninteracting fermions, the partition function can be evaluated
exactly for arbitrary values of the magnetic field:
\egy{Z(\beta,h)=\prod_{k=1}^L (1+e^{-\beta \lambda_k}),}
where $\lambda_k=-\cos(\frac{\pi k}{L+1})-h$ is the one-particle 
spectrum of $\hat H_L$.
\par
Starting out from the partition function one can calculate the 
moments of the magnetization as logarithmic derivatives respect to 
the magnetic field. The expectation value reads:
\egy{\langle \hat M \rangle_{\beta} = \frac 1 \beta 
\frac{\partial \ln Z}{\partial h} - \frac L 2 = 
-\frac 1 2 \sum_{k=1}^L \tanh\left(\frac{\beta\lambda_k}{2}\right),
\label{magncan}}
while the fluctuations can be obtained as:
\egy{\langle \hat M^2 \rangle_{\beta}-\langle \hat M \rangle_{\beta}^2=
\frac{1}{\beta^2} \frac{\partial^2 \ln Z}{\partial h^2}=
\frac 1 4 \sum_{k=1}^L \frac{1}{\cosh^2(\frac{\beta\lambda_k}{2})}.}
\par
\begin{figure}[thb]
\center
\includegraphics[scale=0.55,angle=270]{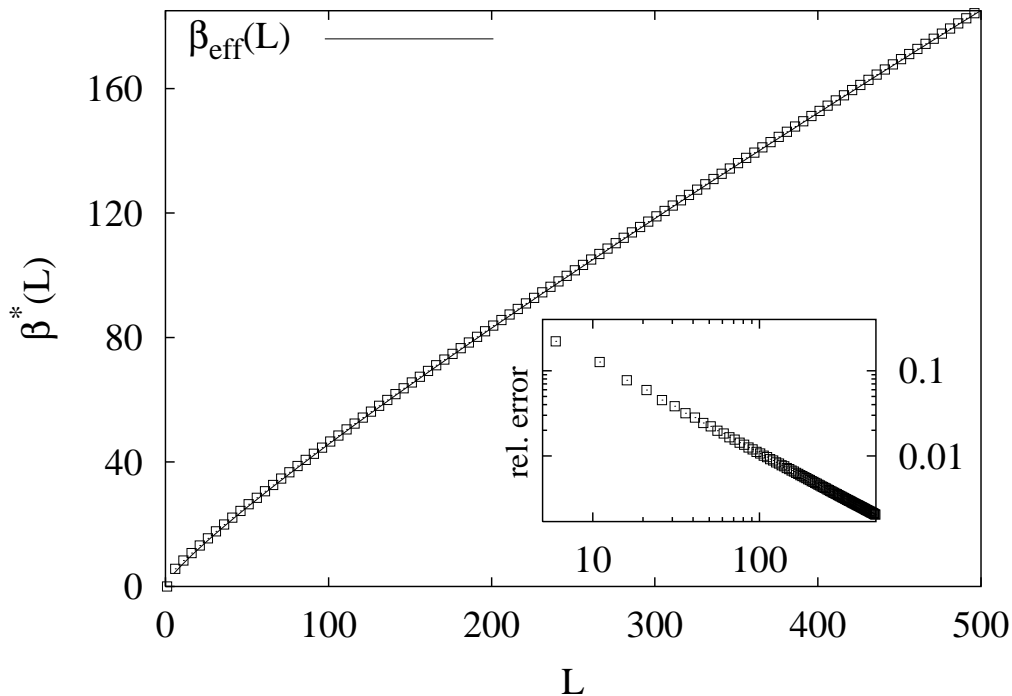}
\caption{Effective inverse temperature obtained by solving
Eq.(\ref{betaeffnum}) numerically (squares). The solid line represents
the asymptotic value Eq.(\ref{betaeff2}) obtained from the 
fluctuation-dissipation relation at zero temperature. 
Inset shows the relative deviation from the asymptotic curve.}
\label{fig:betaeff}
\end{figure}

We have now a closed form for either side of the equation 
(\ref{betaeffnum}), therefore $\beta^*$ can be calculated
numerically for arbitrary system sizes. Figure \ref{fig:betaeff}
shows the results of the calculations together with the
asymptotic form obtained from the fluctuation-dissipation ratio.
%
%
Apparently, either definition for the effective temperature gives the
same result, and the deviation (see inset) from the asymptotic
curve tends to be relatively small even for moderate system sizes.
\par
The precise agreement of the inverse temperatures might look
surprising, but it has a simple explanation in the given case.
Namely it is the consequence of the temperature independent susceptibility.
Indeed, looking at the formula of the magnetization (\ref{magncan}), 
one has a sum of tangent hyperbolic terms, a function which is
known to saturate for large arguments. For large system sizes
$\beta^*$ is expected to be large, thus the nontrivial terms in
the sum are those where $\lambda_k \approx 0$, that is
the Fermi level of the spectrum. However, as already shown in
(\ref{fermilevel}), the spectrum is linear near the Fermi level,
thus application of a small field will cause a shift in the
spectrum, nevertheless preserves linearity. Hence, the terms
in the sum near the Fermi level cancel, the tangent hyperbolic being
an odd function, and the result depends only on the difference in the
number of terms added on the saturated tails. This is of course 
determined by the shift alone that is independent of temperature.
\par
Now, if the susceptibility is temperature independent and we
use the definition (\ref{betafluctdis}) of the effective
temperature, the following relation holds:
\egy{\beta^* \langle \hat M^2 \rangle_{\beta^*}=
\beta_{eff}\langle \hat M_L^2 \rangle,}
which, together with (\ref{betaeffnum}) results in $\beta^*=\beta_{eff}$,
that is the different definitions are identical.
\section{Subsystems in contact \label{sec:contact}}
When one speaks about the temperature of a system there are some
basic properties the definition has to fulfill. For example,
when we take two systems with equal temperatures and place them
in contact, the resulting system must have the same temperature
as the initial ones.
\par
A possible analogue of this situation in our case is the following: we take
two semi-infinite chains with blocks of length $L$ on the boundaries and
join them together. The resulting block of length $2L$ in the infinite
chain has to have an effective temperature dictated by Eq. (\ref{betaeff2}).
\par
In order to solve this problem, we shall calculate the effective inverse
temperature for a block of $L$ spins at the boundary of a semi-infinite chain
using again the fluctuation-dissipation approach for the transverse
magnetization. Following the steps of Section \ref{sec:magfluct} we shall
first calculate the magnetization with a nonzero magnetic field:
\egy{\langle \hat M_L\rangle = 
\sum_{j=1}^{L}\frac 2 \pi \int_0^{q_h}\mathrm{d} q \sin^2(qj)- \frac L 2=
L(\frac{q_h}{\pi}-\frac 1 2)-\frac 1 \pi \sum_{j=1}^{L}
\frac{\sin(2jq_h)}{2j},}
where now brackets denote averages taken in the ground state of
the semi-infinite chain.
\par
One can see that the first term of the above result coincides
with the one obtained for the infinite chain (\ref{eq:maghinf}),
however the second term has an explicit dependence on the location
of the block sites which is the consequence of broken translational symmetry.
Taking the derivative at zero magnetic field, the second term 
becomes zero for even and $-1/\pi$ for odd $L$ thus for even $L$ we
recover the infinite chain result (\ref{magsusc}) exactly.
\par
\begin{figure}[thb]
\center
\includegraphics[scale=0.55,angle=270]{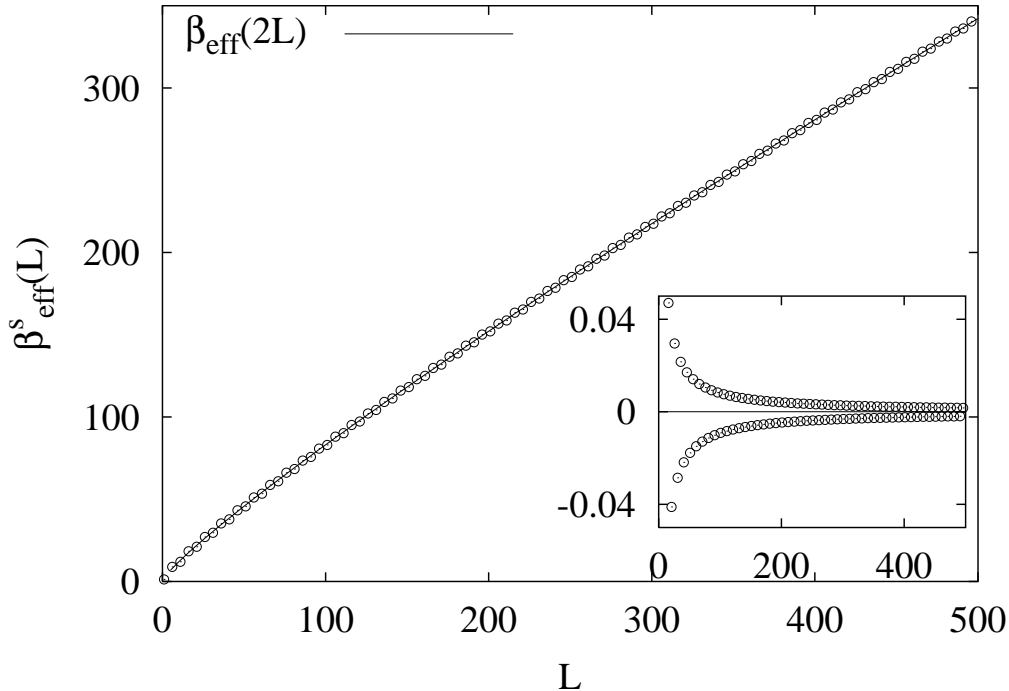}
\caption{Effective inverse temperature of a block of $2L$ spins in an
infinite chain (solid line) and a block of $L$ spins on the boundary
of a semi-infinite chain (circles). The inset shows the relative
deviation from the $\beta_{eff}(2L)$ line. The oscillations are due to
the difference between even/odd block sizes.}
\label{fig:betasemiinf}
\end{figure}

For the fluctuations we need the elements of the correlation matrix
$C_{ij}^s$ of the semi-infinite chain which can be given as:
\egy{C^s_{ij}=\frac{\sin(\frac \pi 2(i-j))}{\pi(i-j)}-
\frac{\sin(\frac \pi 2(i+j))}{\pi(i+j)}.}
Now, one has to simply substitute $C_{ij}$ with $C^s_{ij}$ in 
Eq.(\ref{eq:magfluctex}). Introducing the difference $r=i-j$ and the
sum $s=i+j$ of the sites we have:
\egy{
\langle M_L^2\rangle =
-\frac{2}{\pi^2} \sum_{\substack{r=1 \cr r\, \mathrm{odd}}}^{L-1}
\sum_{\substack{s=2+r \cr s\, \mathrm{odd}}}^{2L-r}
\left[\frac{1}{r^2}+\frac{1}{s^2}+
2\frac{(-1)^\frac{r+s}{2}}{rs}\right] + \frac L 4}
We were able to evaluate the above sum only numerically.
The result for the effective inverse temperature $\beta_{eff}^s(L)$ of a
block of $L$ spins in the semi-infinite chain is shown on Figure
\ref{fig:betasemiinf} together with the effective inverse temperature
$\beta_{eff}(2L)$ of a block of size $2L$ in an infinite chain.
One can see an excellent agreement between the two results along
with oscillations which might originate from the difference between even
and odd block sizes. It is reasonable to think that these oscillations
might be interpreted through the fact that the interaction between
the spins generates anticorrelations between the transverse spin components,
thus all quantities whose definition involves transverse correlations
(such as the effective temperature) will be sensitive to the parity of
the system size.
%
\par
In addition to verifying that the effective temperature does not
change when bringing two subsystems of the same size into contact,
we also obtained the formula
\egy{\beta_{eff}^s=2\pi \frac{L}{\ln L + \ln 2 + c},\label{eq:betasemiinf}}
which is valid when the block is at the boundary of a semi-infinite chain,
and $c$ is the constant defined in (\ref{eq:magfluct}).
This result also confirms the factors close to $2\pi$ instead of $\pi$,
obtained when fitting to the DMRG results with open boundary conditions.
\par
With the formula (\ref{eq:betasemiinf}) in hand, we can also confirm 
another thermodynamic property of the temperature.
Namely, if we take two subsystems of different temperatures and
bring them into contact the resulting temperature of the joint system
must have a temperature between the two initial values.
Changing to effective inverse temperatures, this means that the
following inequalities should hold:
\egy{\beta_{eff}^s(L)<\beta_{eff}(L+L^\prime)<\beta_{eff}^s(L^\prime),
\quad \mathrm{if} \quad L<L^\prime. \label{eq:betaordered}}
In order to prove (\ref{eq:betaordered}), we use the equality
$\beta_{eff}^s(L)=\beta_{eff}(2L)$ and require the positivity
of the derivative of $\beta_{eff}(L+L^\prime)$ respect
to $L$ and $L^\prime$. This requirement yields the condition
$\ln (L+L^\prime)+c>1$. As one can see, this condition holds
even for the smallest possible unequal system sizes
($L=1,\,L^\prime=2$) provided $c>0$ which is what we have found
in all our calculations.
\section{Final remarks \label{sec:fin}}
We have presented an effective thermodynamic description
of an entangled block in an XX spin chain at zero temperature.
For a better understanding of the range of validity of this picture
several comments are in order.
\par
First of all, it has to be noted that during our investigation
of the one-particle spectrum of the effective Hamiltonian
(\ref{effhamspect}), we dealt only with the eigenvalues and did not
look at the appropriate eigenstates. It was already demonstrated in
\cite{PeschelXX} that, in contrast to the homogeneous plane wave 
eigenstates of the open XX chain, the eigenstates of the embedded block
show enhanced amplitudes near the boundaries. Nevertheless, all
those states with symmetrically occupied particle-hole excitations,
such as the ground state, are identical.
Furthermore, certain quantities such as the transverse magnetization
have the same expectation values in either of the one-particle states.
However, it is clear that the role of this strange amplitude enhancement
requires deeper understanding, especially when dealing with quantities
that are sensitive to boundary effects. It is possible that
in such cases the validity of this simple thermodynamic picture
might be lost, or at least requires more careful considerations.
\par
It would be also desirable to test the fluctuation-dissipation
approach with some nonconserved quantity e.g. the longitudinal
magnetization. Unfortunately, in such cases the calculation of
the response function is fairly involved, since the model
becomes nonintegrable when the appropriate field is switched on.
Even so, there remains the possibility
to determine the response function numerically by means of
DMRG calculations for finite chains.
\par
It is also reasonable to expect, that the effective temperature description
might be carried over to other simple conformaly invariant models,
such as the critical transverse Ising chain, where the form of the
reduced density matrix of a subsystem is rather similar to the XX case
\cite{Legeza,Orus}, however the exact form of the effective temperature
might be somewhat different.
\par
One can also speculate on the effect of placing the system in a heat
bath of finite inverse temperature $\beta_{hb}$. It is reasonable to
expect that the effective temperature should not change dramatically
as long as thermal fluctuations are subdominant of the quantum fluctuations, 
i.e. when $\beta_{hb}>>\beta_{eff}$.
On the other hand, when the temperature of the heat bath becomes
much larger then the effective temperature ($\beta_{hb}<<\beta_{eff}$)
the subsystem should be thermalized by the heat bath. Therefore,
the behaviour of a subsystem should change, when the inverse temperature
of the heat bath is on a scale $\beta_{hb}\propto \frac{L}{\ln L}$.
This condition has been actually found for the transition temperature
between the logarithmic and extensive behaviour of the entropy in
gapless free fermionic systems \cite{Gioev}, further supporting the
existence of an effective temperature.
\par
As far as higher dimensional systems are considered, we remark on
an other result of \cite{Gioev} where the fluctuations of the
transverse magnetization has been found to scale as $L^{d-1}\ln L$
with the system size in arbitrary dimensions. This property together
with the extensive character of the susceptibility might lead us to
speculate, that our fluctuation-dissipation approach would result
an effective temperature with a same dependence on the linear size
also in higher dimensions. We emphasize however, that these 
arguments require more careful investigations.
\par
Finally, it should be pointed out that there is an opportunity
of extending our calculations to nonequilibrium situations,
such as the driven XX chain. We believe, that the simple definitions
we applied in the equilibrium case could be easily carried over,
and the results may shed light on the possibility of introducing an
effective temperature in systems far from equilibrium.

\ack We would like to thank Ingo Peschel for a stimulating discussion.
This work was financially supported by OTKA Grants No. T043734, TS044839,
T043330 and F046356 and the J\'anos Bolyai Research Fund.

\end{document}